\documentclass[prl,showpacs,preprintnumbers,twocolumn,amsmath,amssymb,superscriptaddress]{revtex4-1}
\usepackage{graphicx}
\usepackage{dcolumn}
\usepackage{bm}
\usepackage{psfrag}
\usepackage[normalem]{ulem}
\usepackage{cancel}

\usepackage{color}
\DeclareMathOperator{\sgn}{sgn}
\DeclareMathOperator{\diverg}{div}

\begin{document}
\title{Intrinsic spin Hall effect in systems with striped spin-orbit coupling}
\author{G\"otz Seibold}
\affiliation{Institut f\"ur Physik, BTU Cottbus-Senftenberg, PBox 101344, 03013 Cottbus, Germany}
\author{Sergio Caprara}
\affiliation{Dipartimento di Fisica - Universit\`a di Roma Sapienza, piazzale Aldo Moro 5, I-00185 Roma, Italy}
\author{Marco Grilli}
\affiliation{Dipartimento di Fisica - Universit\`a di Roma Sapienza, piazzale Aldo Moro 5, I-00185 Roma, Italy}
\author{Roberto Raimondi}
\affiliation{Dipartimento di Matematica e Fisica, Universit\`a Roma Tre, Via della Vasca Navale 84, 00146 Rome, Italy}
\begin{abstract}
The Rashba spin-orbit coupling arising from structure inversion asymmetry couples spin and momentum degrees of freedom providing 
a suitable (and very intensively investigated) environment for spintronic effects and devices. Here we show that in the presence 
of strong disorder, non-homogeneity in the spin-orbit coupling gives rise to a finite spin Hall conductivity in contrast with 
the corresponding case of a homogeneous linear spin-orbit coupling. 
 In particular, we examine the inhomogeneity arising from a striped structure for a 
two-dimensional electron gas, affecting both density and Rashba spin-orbit coupling. We suggest that this situation
can be realized at oxide interfaces with periodic top gating.
\end{abstract}
\maketitle

\begin{figure}[htb]
\includegraphics[width=6cm,clip=true]{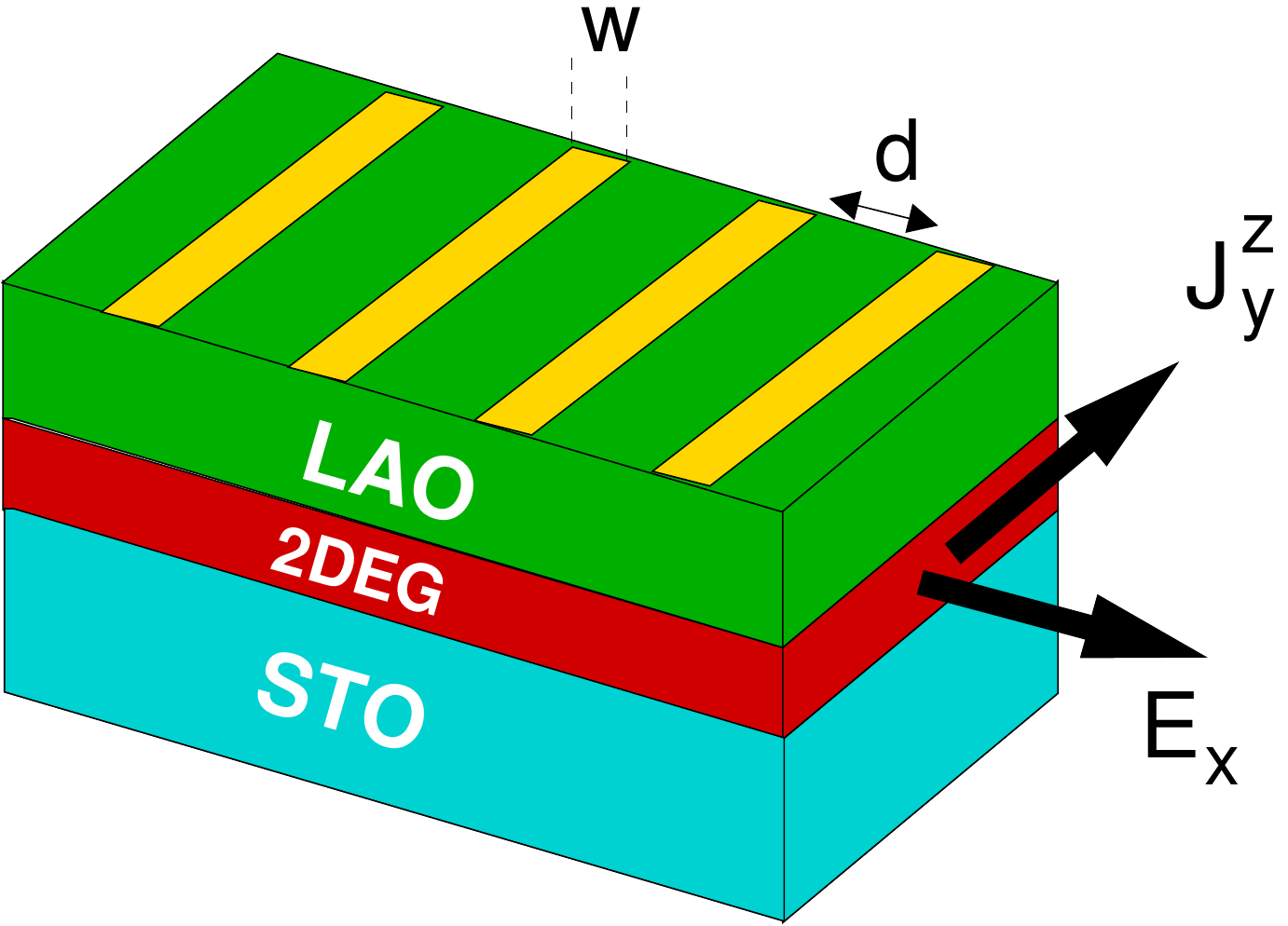}
\caption{Schematic view of a possible device in which the SHE is enforced in the 2DEG at the interface of a LAO/STO 
heterostructure. The yellow stripes represent top-gating electrodes of width $w$ and interspacing $d$.}
\label{figSHE1}
\end{figure}

The spin Hall effect (SHE) \cite{dyakonov1971} is the generation of a transverse spin current by an applied electric field with 
the current spin polarization being perpendicular to both the field and the current flow. Since the SHE allows the control of the 
spin degrees of freedom even without external magnetic fields 
(see e.g. Refs. \cite{vignale2010,jungwirth2012}), it has become a central topic in present spintronics research.~\cite{zutic04,maekawa13} 
The microscopic origin of the SHE lies in the spin-orbit coupling (SOC), which in solid-state systems may be due to the potential 
of the ionic cores of the host lattice, the potential of the impurities and the confinement potential of the device structure. In 
a two-dimensional electron gas (2DEG), Bychkov and Rashba \cite{bychkov1984} have proposed that the lack of inversion symmetry 
along the direction perpendicular to the gas plane leads to a momentum-dependent spin splitting usually described by the 
so-called Rashba Hamiltonian
\begin{equation}
\label{rashba_ham}
H=\frac{p^2}{2m}+\alpha {\boldsymbol\tau}\times {\bf z}\cdot {\bf p},
\end{equation}
where ${\bf p}$ is the momentum operator for the motion along the plane hosting the 2DEG, say the xy plane, ${\bf z}$ is a 
unit vector perpendicular to it, ${\boldsymbol \tau}=(\tau^x,\tau^y,\tau^z)$ is the vector of the Pauli matrices, and $\alpha$ is
a coupling constant whose strength depends on both the SOC of the material and the field responsible for the parity breaking. 
The Hamiltonian (\ref{rashba_ham}), which has been extensively used in the study of the 2DEG in semiconducting systems, has been 
recently applied also to interface states between different metals \cite{roja-sanchez2013} and between two insulating oxides
\cite{caviglia2010,fete2012,hayden1913,bergeal2015}. In the latter systems, higher mobilities, carrier concentration and SOC strengths have 
led to the expectation of observing stronger SOC-induced effects. The Hamiltonian (\ref{rashba_ham}) is deceptively simple, as 
one realizes when considering transport phenomena. In particular, the intrinsic universal SHE proposed in Ref.\,\cite{sin04} 
turned out to be a non stationary effect, while under stationary conditions cancellations occur, leading to a vanishing spin 
Hall conductivity (SHC) $\sigma^{sH}$, i.e., the coefficient relating the $z$-spin current in the $y$ direction to 
the applied electric field, $J^z_y=\sigma^{sH} E_x$.
Here, we show that this is only true for a spatially homogeneous $\alpha$, while considering a space-dependent 
$\alpha(x,y)$ opens the way to a substantial SHE under stationary conditions, even in the presence of strong disorder. We shall 
first discuss from a general perspective how this comes about, and shall afterwards demonstrate numerically the effect in the 
presence of a spatially modulated SOC as it could be realized in the 2DEG at the interface of
a LaAlO$_3$/SrTiO$_3$ (LAO/STO) heterostructure, schematically depicted in Fig.\,\ref{figSHE1}.
It is experimentally established that the Rashba SOC increases when the electron density in the 2DEG of these heterostructures 
is increased \cite{nitta97,caviglia2010,fete2012,bergeal2015}. Since the local electric field determining $\alpha$ is tightly related 
to the electron density \cite{marco12}, one can naturally infer that the structure in Fig.\,\ref{figSHE1} produces a modulation 
of the Rashba SOC.

{\it --- General arguments --- }
The interplay of the {\it intrinsic} Rashba SOC with the scattering from impurities makes the dynamics of charge and spin degrees 
of freedom intrinsically coupled in a subtle way. This is especially evident in the vanishing of the SHC with homogeneity in the SOC. Notice that the spin 
current is a tensor quantity depending on the flow direction (lower index) and spin polarization axis (upper index) and hence the 
spin current and the electric field are related by a tensor of third rank $\sigma^a_{ij}$. For fixed polarization $a=z$, 
Onsager's relations require the antisymmetry property $\sigma^z_{yx}=-\sigma^z_{xy}=\sigma^{sH}$. 

The vanishing of the SHC manifests via an exact compensation of the contribution originally proposed by Sinova et al. \cite{sin04} 
by a further contribution, which arises by the coupling between the spin current and the spin polarization, which is induced in 
the plane perpendicularly to the applied electric field. This latter effect was almost simultaneously proposed by Edelstein 
\cite{edel} and by Aronov and Lyanda-Geller \cite{aronov89}. It consists of a non-equilibrium spin polarization due to the 
electric field $S^y_0=-e \alpha N_0 \tau E_x$ for a field $E_x$ along the $x$ axis. Here $e$ is the unit charge ($-e$ for 
electrons), $N_0=m/2\pi\hbar^2$ is the density of states per spin of the 2DEG described by the first term of Eq. (\ref{rashba_ham}) and $\tau$ is 
the elastic relaxation time introduced by impurity scattering. In such a 2DEG the standard Drude formula can be written via 
the Einstein relation as $\sigma =2e^2 N_0 D$, with the diffusion coefficient $D=v_F^2 \tau/2$, $v_F$ being the Fermi velocity 
related to the Fermi energy $E_F=m v_F^2/2$. To understand the origin of the compensation mentioned above, it is useful to start 
from a property of the Hamiltonian (\ref{rashba_ham}) first pointed out by Dimitrova \cite{dimitrova05}, which relates the 
time derivative of the $S^y$ spin polarization to the spin current
\begin{equation}
\label{dimitrova}
\partial_t S^y=-2m\alpha\hbar^{-1} J^z_y.
\end{equation}
Notice that such a relation is not changed by disorder scattering as long as the latter is spin independent. 
Disorder is necessary to guarantee a stationary state which implies the left-hand side of Eq. (\ref{dimitrova}) to be zero.
Obviously, the corresponding vanishing of the right-hand side entails a vanishing SHC when $\alpha$ is a constant. The specific 
way in which the vanishing of the SHC occurs in a disordered 2DEG via the so-called vertex corrections
\cite{inoue2004,mishchenko2004,raimondi2005,khaetskii2006} can be heuristically understood by describing the coupling between 
spin current and spin polarization as a generalized diffusion in spin space. By dimensional arguments spin current and spin 
density must be related by the factor $L_{SO}/\tau_{DP}$, where $L_{SO}=\hbar /(2m\alpha)$ is the spin-orbit precession 
length originating from the difference of the Fermi momenta of the two branches of the spectrum 
of Eq. (\ref{rashba_ham}), while  $\tau_{DP}$ is the Dyakonov-Perel spin relaxation time due to the interplay of SOC and 
disorder scattering. In a disordered 2DEG, $\tau_{DP}$ is related to $L_{SO}$ by the diffusion coefficient $L_{SO}^2=D\tau_{DP}$, 
so that one obtains for the spin current
\begin{equation}
\label{spin_current}
J^z_y=2m\alpha\hbar^{-1} D S^y +\sigma^{sH}_0 E_x.
\end{equation}
In the above equation, which can be rigorously derived \cite{gorini2010,raimondi2012}, the quantity $\sigma^{sH}_0$ is the 
intrinsic contribution of Ref.\,\cite{sin04} in the diffusive regime $\alpha p_F \tau /\hbar \ll 1$, whereas the term proportional to $S^y$ corresponds to the vertex-correction 
contribution mentioned above. Given the expression for $\sigma_0^{sH}= (e/8\pi) (2\tau /\tau_{DP})$ derived in 
Ref.\,\onlinecite{sin04} it is now apparent that, if we replace $S^y$ with the Edelstein result, the spin current 
in Eq. (\ref{spin_current}) vanishes, consistently with the stationarity requirement derived from Eq. (\ref{dimitrova}). The 
key observation is that this compensation does not necessarily occur for an inhomogeneous SOC, where it is no longer possible to
express the time derivative of the spin polarization in terms of the spin current. In such a situation a spin current becomes 
possible under stationary conditions. 

In order to illustrate the physical mechanism by which an inhomogeneous Rashba SOC leads to a finite SHE
it is useful to consider a single-interface problem, which is described by Eq. (\ref{rashba_ham}) 
with the replacement $\alpha \rightarrow \alpha (x)=\theta (x)\alpha_++\theta (-x)\alpha_-$ (with $\alpha_+>\alpha_-$).
Clearly as $x\to\pm \infty$, one recovers the uniform case with complete cancellation of the spin current for the Rashba model 
with couplings $\alpha_{\pm}$. On both sides of the interface, the $y$-spin polarization obeys a diffusion-like equation with  
$L_{\pm}$, the corresponding spin-orbit lengths in the two regions. One can then seek a solution of the form
\begin{eqnarray*}
S^y(x)&=&\theta (x)\left( S_{0,+}+\delta s_+ e^{-x/L_+}\right)\nonumber\\&+&
\theta (-x)\left( S_{0,-}+\delta s_- e^{x/L_+}\right),\nonumber
\end{eqnarray*}
where $S_{0,\pm}$ are the asymptotic values of the $y$-spin polarization at $\pm\infty$. The constants $\delta s_{\pm}$
must be determined by imposing the appropriate boundary conditions at $x=0$. As a result, the spin current is exponentially 
localized near the interface, where the spin polarization $S^y(x)$ must interpolate between the two asymptotic values and there 
is no longer complete cancellation between the two terms of Eq. (\ref{spin_current}). 

One can imagine to generalize this analysis to a series of interfaces apt to describe a periodic modulation of the SOC. Expectedly, 
if the spin-orbit length is larger than the distance between two successive interfaces, the spin Hall current should become 
practically uniform. 

{\it --- The model and its numerical solution ---}
The previous arguments within the diffusive limit are
now substantiated by numerical results for a microscopic 2D lattice model (size $N_x\times N_y$) with inhomogeneous Rashba 
SOC described by the Hamiltonian
\begin{equation}                                         
H=\sum_{ij\sigma}t_{ij}c^\dagger_{i\sigma}c_{j\sigma} 
+ \sum_{i\sigma}(V_i-\mu) c^\dagger_{i\sigma}c_{i\sigma} + H^{RSO},  \label{eq:ham}
\end{equation}   
where $c^\dagger_{i\sigma}(c_{i\sigma})$ creates (annihilates) an electron with spin projection $\sigma$ on the site
identified by the lattice vector $R_i$, the first term describes the kinetic energy of electrons on a 
square lattice (lattice constant $a$, only nearest-neighbor hopping:
$t_{ij}\equiv -t$ for $|R_i-R_j|=a$) and in the second term $V_i$ is a
local disorder potential with a flat distribution $-V_0 \le V_i \le V_0$, and $\mu$ is the chemical potential.
The last term is the lattice Rashba SOC,
\begin{eqnarray*}
H^{RSO}&=&-i\sum_{i\sigma\sigma'} \alpha_{i,i+x} 
\left\lbrack c^\dagger_{i\sigma} \tau^y_{\sigma\sigma'} c_{i+x,\sigma'} 
- c^\dagger_{i+x,\sigma} \tau^y_{\sigma'\sigma} c_{i,\sigma} \right\rbrack\nonumber\\ &+& i \sum_{i\sigma\sigma'} \alpha_{i,i+y}
\left\lbrack c^\dagger_{i\sigma} \tau^x_{\sigma\sigma'} c_{i+y,\sigma'} 
- c^\dagger_{i+y,\sigma} \tau^x_{\sigma'\sigma} c_{i,\sigma} \right\rbrack\nonumber
\end{eqnarray*}
and the coupling constants $\alpha_{i,i+x/y}>0$ are now defined on the bonds. Note that for constant 
$\alpha\equiv\alpha_{i,i+x/y}$ Eq. (\ref{eq:ham}) takes the usual form [2nd term in Eq. (\ref{rashba_ham})] 
in momentum space \cite{goetz15}.

One can show \cite{SM} that a ``continuity equation'' for the local spin density $S_i^y$, (a dot stands for time derivative)
\begin{equation}\label{eq:relation}
{\dot S_i^y}+[\diverg {\bf J}^y]_i + \alpha_{i,i+y} J^z_{i,i+y} + \alpha_{i-y,i} J^z_{i-y,i} = 0,
\end{equation}
holds. For a homogeneous Rashba SOC, where $[\diverg {\bf J}^y]_i =0$, Eq. (\ref{eq:relation}) corresponds to 
Eq. (\ref{dimitrova}) and implies that the total $z$-spin current has to vanish under stationary conditions.  
On the contrary, when $\alpha$ varies in space, a cancellation occurs \cite{SM} between $\diverg J^y$ and the last two terms 
of Eq. (\ref{eq:relation}), so that the stationarity condition $\dot S=0$ can be fulfilled without the vanishing of 
$J^z$. This is also clear for a system with periodic boundary conditions, where the total divergence of any current has to 
vanish, i.e., 
\begin{equation}\label{eq:sumtot}
-\sum_i {\dot S_i^y}= \sum_i \left\lbrace \alpha_{i,i+y} J^z_{i,i+y} + \alpha_{i-y,
i} J^z_{i-y,i} \right\rbrace .
\end{equation}
Clearly, the left-hand side can vanish without implying $J^z=0$, because, if $\alpha$ is inhomogeneous,
Eq. (\ref{eq:sumtot}) can be fulfilled with alternating signs of $J^z$ in regions with different $\alpha$ [see the top panel of Fig. 4 in Ref. \onlinecite{SM}].

\begin{figure}[htb]
\includegraphics[width=8.5cm,clip=true]{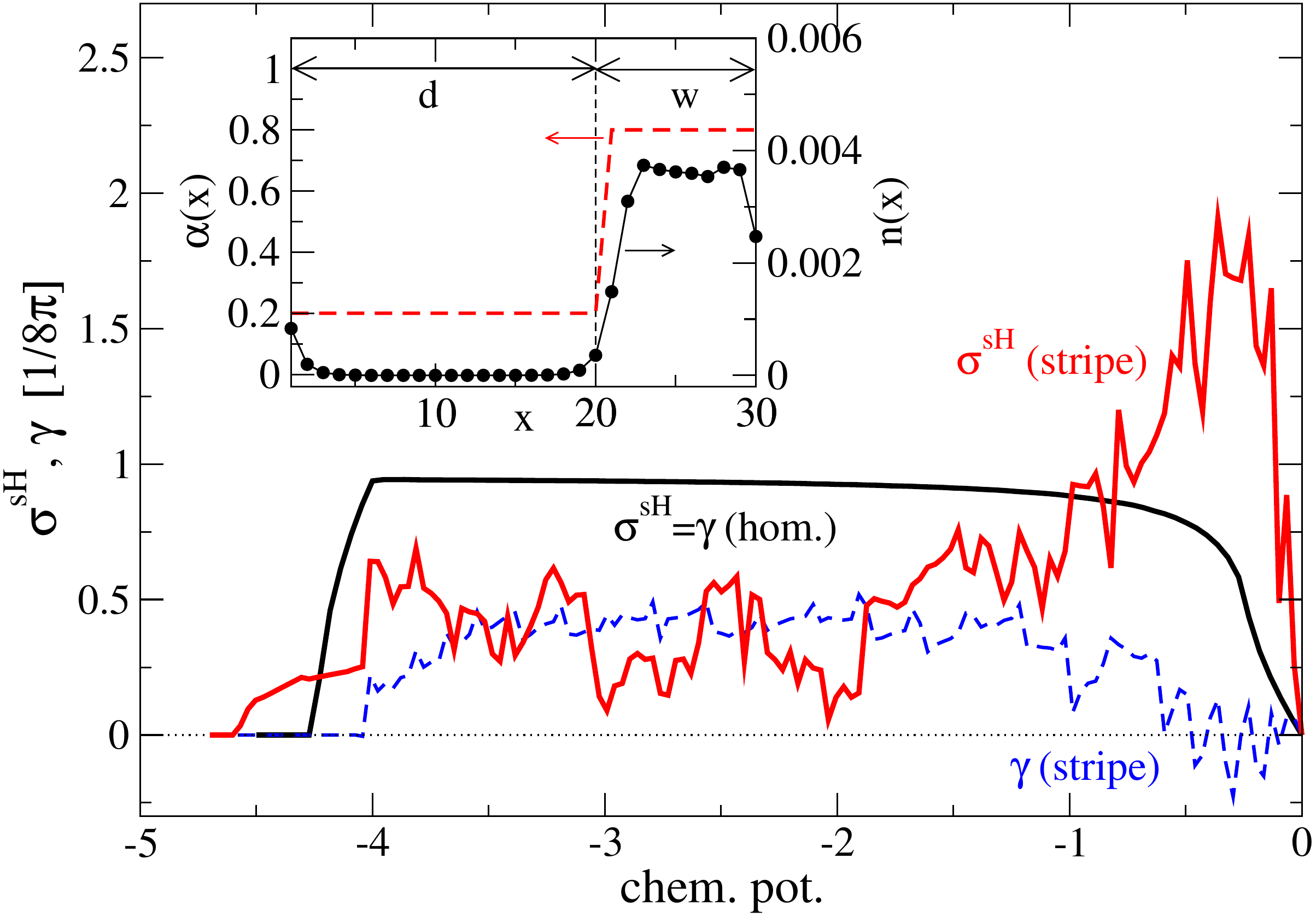}
\caption{Main panel: Spin Hall coefficient $\sigma^{sH}$ and ``stationarity''
parameter $\gamma$ [both in units of $1/(8\pi)$] as a function of $\mu$ 
for a homogeneous system with $\alpha=0.5\, t$ (black solid line) 
and a striped system ($\sigma^{sH}$: red solid and $\gamma$: blue dashed) 
with modulated Rashba SOC as shown in the inset.
Here the (red) dashed line displays
the variation of $\alpha(x)\equiv \alpha_{i,i+x}=\alpha_{i,i+y}$ 
along the $x$ direction for stripes along the $y$ direction and 
width $w=10\, a$ separated by a distance $d=20\, a$. The Rashba SOC on the stripes
is $\alpha(x)\equiv \alpha_1=0.8\,t$ while between the stripes 
$\alpha(x)\equiv \alpha_0=0.2\,t$.
The black solid line (dots) in the inset reports the charge profile
at chemical potential $\mu=-4.3\,t$.  
System size: $3060\times 3060$ sites.}
\label{figSHE2}
\end{figure}
\begin{figure}[htb]
\includegraphics[width=8.5cm,clip=true]{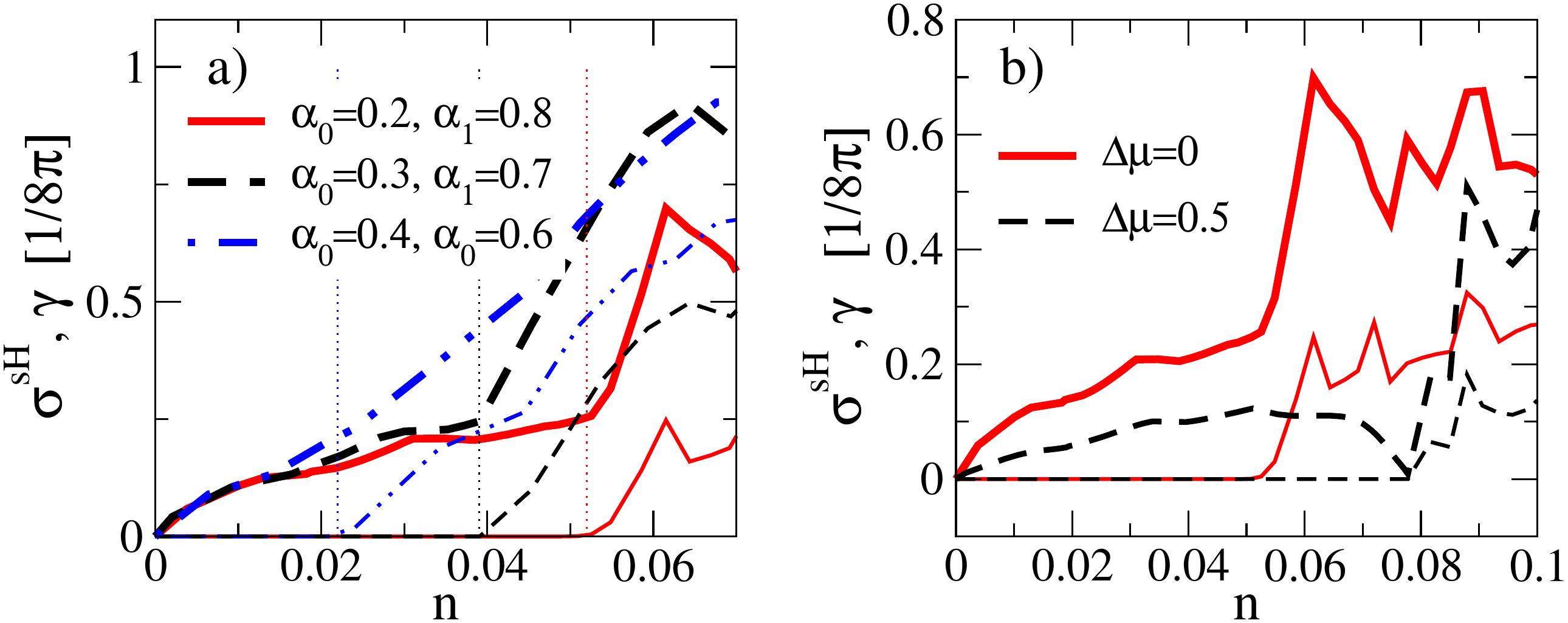}
\caption{
a): $\sigma^{sH}$ (thick lines) and $\gamma$ (thin lines) as a function of density $n$ (average number of electrons per lattice 
site) for a striped system and parameters are indicated in the panel.
b): $\sigma^{sH}$ (thick lines) and $\gamma$ (thin lines) as a function of density but now with an 
additional modulation of the local chemical potential $\mu^{loc}_{ix}$ which
is set to $-0.5\, t$ on the stripes ($a_0=0.2\, t$, $a_1=0.8\, t$). }
\label{figSHE3}
\end{figure}

We exemplify the situation for an inhomogeneous Rashba SOC which varies along the $x$ direction
forming a superlattice with $d=20a$ and $w=10a$ [see Fig.\,\ref{figSHE1}]. In the regions of width $d$,
$\alpha=a_0$ is smaller than in the regions of width $w$, where $\alpha=a_1>a_0$.
The inhomogeneity in $\alpha_{i,i+x,y}$ leads to a concomitant charge
modulation which is shown in Fig.\,\ref{figSHE2}(a).
We have diagonalized the Hamiltonian (\ref{eq:ham}) and calculated the SHC from the Kubo formula 
(see, e.g., Ref.\,\onlinecite{sin04}) $\sigma^{sH}=\sum_{ij}\sigma^{sH}_{ij}$ with 
\begin{equation}
\sigma^{sH}_{ij}\equiv  \frac{2}{N}\sum_{\substack{E_n < E_F\\E_m > E_F}}
\frac{Im \langle n|j^z_{i,i+y}|m\rangle
\langle m|j^{ch}_{j,j+x}|n\rangle}{(E_n-E_m)^2+\eta^2} \,.\label{eq:kubo1}
\end{equation}
Here we have taken the limit of zero temperature and $\eta \to 0$ is a small regularization term which can be interpreted 
as an inverse electric-field turn-on time \cite{nomura05} . Note that for the striped system one has already spin currents 
$J^{z,0}_{i,i+y}$ flowing in the ground state \cite{sonin07} (see, e.g., Fig.\,3 in Ref.\,\onlinecite{SM}) whereas the electrically induced 
spin current is $J^{z,ind}_{i,i+y}=\sum_j\sigma^{sH}_{ij} E_{x}$. Thus Eq. (\ref{eq:sumtot}) can be split into a ground-state 
contribution (for which of course ${\dot S_i^y}=0$) and a linear-response part. For the latter we define the quantity
$\gamma=2\sum_{ij}\alpha_{i,i+y}\sigma_{ij}^{sH}$ which therefore also describes the linear response of 
$\sum_i\dot{S_i^y}=-\gamma E_x$ to the applied electric field and which in the following will be used to quantify the 
``stationarity'' of the solution. In fact, Fig.\,\ref{figSHE2} demonstrates that for a constant $\alpha=0.5\,t$ 
(black solid line) $\sigma^{sH}$ coincides with $\gamma$ and therefore the finite $\sigma_{sH}$ is a non-stationary result.
On the other hand, the same panel also reports the results for the case $a_0=0.2\,t$ and $a_1=0.8\,t$. In this case one can 
see that for a non-negligible range of chemical potential near the bottom of the band a substantial $\sigma^{sH}$ (red solid curve) 
is present while $\gamma=0$ (blue dashed curve), marking the occurrence of a SHE in stationary conditions. 
This indicates the relevant role of those states that are still extended along the $y$ direction, while they 
are nearly localized inside the potential wells arising from the modulation of $\alpha$.
As shown in Fig.\,\ref{figSHE3}(a), this situation occurs for increasingly large density ranges by increasing the inhomogeneity of 
$\alpha$. We have also checked that the stationarity is not only global but that in the low density regime 
${\dot S_i} \approx 0$ is fulfilled at each lattice site.

We also investigated the effect of an inhomogeneous chemical potential as it is induced by the striped gating of 
Fig.\,\ref{figSHE1}. In particular, we shift the chemical potential downwards by $\Delta \mu=0.5\,t$ on the sites below the gate 
(the regions of width $w$ with $\alpha=a_1=0.8\,t$. From Fig.\,\ref{figSHE3}(b) one sees that, although $ \sigma^{sH}$ is reduced, 
it still remains substantial and the density range with a stationary SHE is even extended (black dashed curves). On the contrary 
one can see \cite{SM} that, in the absence of an inhomogeneous Rashba SOC, a simple charge modulation does not produce any 
stationary SHE.

\begin{figure}[htb]
\includegraphics[width=8cm,clip=true]{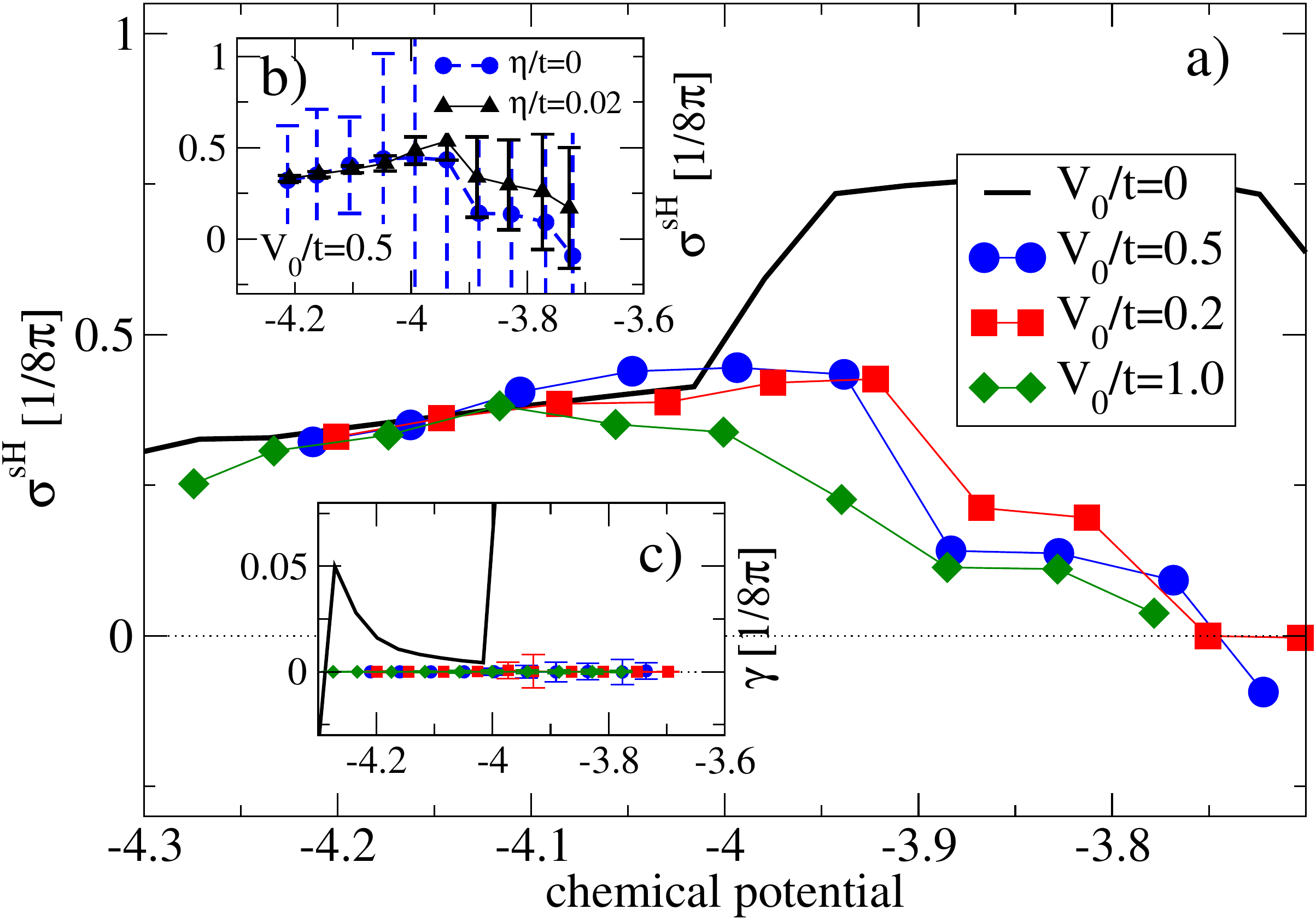}
\caption{Main panel (a): Spin Hall conductivity as a function of 
chemical potential $\mu$
and various values of disorder. Panel (b): $\sigma^{sH}$
as a function of
$\mu$ including error bars for $V_0/t=0.5$ and two values of 
$\eta/t=0$ (circles) and $\eta/t=0.02$ (triangles).  
Panel (c) reports the behavior
of the ``stationarity'' parameter $\gamma$. Results in panels (a,c) are
obtained for $\eta=0$.}
\label{figSHE4}
\end{figure}

Finally, we address the quite important issue of the robustness of SHE in the presence of disorder. Previous analyses 
\cite{sheng05,nomura05} showed that the SHC for a linear Rashba SOC is rapidly destroyed by disorder. This is easily understood 
because in the homogeneous Rashba SOC the SHE is a non-stationary effect which cannot survive the relaxation effect of 
disorder scattering. Here, instead, when $\gamma=0$ the SHE is present in a stationary state and disorder is much less 
effective in spoiling it. In the presence of a random potential of finite width $V_0$ the calculations can only be carried 
out on smaller lattices ($40\times 40$ sites) where we consider stripes of width $w=4\,a$ and
distance $d=4\,a$ (see Fig. \ref{figSHE1}). We follow the procedure described in Ref.\,\onlinecite{sheng05} 
and diagonalize the Hamiltonian for different disorder configurations and different twisted boundary conditions.
For each concentration we consider $250$ random boundary phases and $50$ disorder 
configurations. 

As a striking result (main panel of Fig. \ref{figSHE4}) 
we find that the average SHC at low densities ($\mu < -4\,t$) is 
not affected by disorder and only gets suppressed when the chemical potential is within the range of band states extended 
both along $x$ and $y$ directions. In contrast, and as mentioned above, $ \sigma^{sH}$ vanishes for a 
homogeneous, linear Rashba SOC \cite{sheng05,nomura05} in the presence of disorder and for $\eta \to 0$.

It has been pointed out in Ref. \cite{nomura05} that
the evaluation of $ \sigma^{sH}$ on finite lattices and taking
the limit $\eta \to 0$ is complicated by strong fluctuations. These 
strong variances in the SHC are exemplified in
panel (b) of  Fig. \ref{figSHE4} for $V_0/t=0.5$ where we also 
show the corresponding result for $\eta/t=0.02$. The SHC in the
low density regime is not dependent on the small $\eta$ value but
one observes a large reduction in the variance which becomes of the
order of the symbol size. We therefore can safely conclude
that our finite size results support a finite SHC at low densities
even for strongly disordered systems. 

Naturally, the system gets more stationary with disorder, as it is shown in panel (c) of Fig.\,\ref{figSHE4}, where 
a small residual value of $\gamma$ for the clean striped system (black curve) is suppressed for all $\mu$'s. 
Again, for $\mu<-4\,t$ this does not imply the vanishing of $\sigma^{sH}$, as would be the case for homogeneous Rashba SOC.

{\it --- Discussion and conclusions ---} 
The above analysis clearly demonstrates that a system with modulated Rashba SOC can sustain a finite
SHE in stationary conditions. This occurs for a limited density range, when the chemical potential falls in a region 
where the states are strongly affected by the modulated $\alpha(x)$ and are almost localized in the bottom of the 
modulating potential (in the direction of the modulation; the states are extended in the perpendicular direction). 
Therefore the response of the charge current $J_x^{ch}$ to the electric field along the modulation direction is strongly 
suppressed which can lead to large spin Hall angles $e J^z_y/J_x^{ch}$ for the striped system. It is important to note that 
this effect is due to the modulation of the Rashba SOC and cannot arise in a ``conventional'' charge density wave. In fact, 
for constant $\alpha$ and independent of the electronic structure Eq. (\ref{eq:sumtot}) predicts the vanishing
of the SHE under stationary conditions. 

The implementation of this analysis in a real system is for sure a challenging task for several reasons. First of all the top 
gating structuring has to be sharp enough to produce a sufficiently sharp spatial modulation of the 2DEG  below the LAO layer 
(which is at least 20\,nm thick): if the modulation of the SOC is not sharp enough on the $L_{SO}$ scale, the 2DEG would feel 
a nearly uniform $\alpha$ and the SHE is expected to vanish. One should also consider that our analysis is based on a simple 
one-band model, while the SOC in the 2DEG in the LAO/STO involves several bands \cite{zhong,khalsa,marco12}. Of course, the 
basic ideas of this work could be tested and hopefully implemented in other, perhaps simpler, systems involving heterostructures 
of semiconductors with modulated Rashba SOC which have been already discussed in the literature in different contexts 
\cite{wang04,japa09,malard11}.

We acknowledge interesting discussions with N. Bergeal, V. Brosco, and J. Lesueur.
G.S. acknowledges support from the Deutsche Forschungsgemeinschaft. 
M.G. and S.C. acknowledge financial support from   Sapienza University of Rome, project Awards
No. C26H13KZS9.

\begin{appendix}
\vspace{2 truecm}
\centerline{\bf APPENDIX }
\section{Rashba systems in the diffusive limit}
To gain insight on the numerical results discussed in the main text, we discuss here a continuum Rashba model in the diffusive limit.
The corresponding diffusion equations can be derived from a microscopic model by using e.g. the Keldysh technique. 
Such a derivation is, for instance given in Ref. \onlinecite{raimondi2006}. For the following discussion, however,
one does not need to know such a microscopic derivation in detail. One main advantage of the diffusion equation 
description is that it contains all the important aspects of the Rashba model and allows an almost analytic treatment, which helps in elucidating the physics.
We first provide the diffusion equation description 
for the uniform case. This is very standard and a recent discussion can be found in Ref. \onlinecite{raimondi2012}.
Then we describe the single-interface problem as the simplest realization  of a  non-uniform Rashba system with two regions
with different Rashba SOC. A subsequent subsection reports the two-interface problem.

\subsection{The uniform case}
In the presence of the Rashba spin-orbit coupling, the spin polarization along the y direction obeys the
 following equation in the diffusive regime (see Ref. \onlinecite{raimondi2006} for a derivation)
\begin{equation}
-\partial_t s^y+D \partial_x^2 s^y=\frac{1}{\tau_{DP}}(s^y -s_0),
\label{eq1}
\end{equation}
where $D=v_F^2\tau/2$ is the standard diffusion coefficient.
$\tau_{DP}^{-1}= (2m\alpha)^2 D$ is the inverse Dyakonov-Perel spin relaxation time. $s_0=-e N_0 \alpha \tau E_x$ is the non-equilibrium spin polarization induced by the electric field $E_x$ applied along the x direction. 
Such a non-equilibrium polarization is sometimes called the Edelstein effect or the spin-galvanic effect.
Here $N_0=m/2\pi\hbar^2$ is the 2D density of states and, in the following, we take units such that $\hbar =1$.
We consider only the dependence on the x direction in the diffusion equation to make contact with the numerical calculation.
 In stationary and uniform circumstances, we must have $s^y=s_0$.
This leads to the vanishing of the spin current $J^z_y$, as it is well known in the Rashba model (see for instance Ref. \onlinecite{raimondi2012}).
To see this, consider that the spin current is given by
two terms: a "drift-like" Hall term and a "diffusion-like" one.
The drift term corresponds to the  calculation of Ref. \onlinecite{sin04}, i.e. it is just the Drude formula for the spin Hall conductivity.
As for the ordinary Hall conductivity, it is non-zero and finite even in the absence of disorder. The diffusion term is usually expected to vanish in uniform situations. However as derived in Ref. \onlinecite{gorini2010} and used in Ref.  \onlinecite{raimondi2012}, the Rashba interaction can be described in terms of a SU(2) vector potential, which then introduces covariant derivatives. The latter are defined by
\begin{equation}
\label{covariant}
\left[ \tilde \nabla_i s\right]^a=\nabla_i s^a -\epsilon^{abc} A^b_i s^c
\end{equation}
where $A^a_i$ is the SU(2) vector potential. In the Rashba case, $A^x_y=-A^y_x=2m \alpha$. The key observation is that the diffusion term related to the covariant derivative of the spin density is present even in the uniform case.
Specifically the spin Hall current we are interested in reads
\begin{equation}
J^z_y=\sigma_0^{sH} E_x +\frac{D}{L_{SO}} s^y,
\label{eq2}
\end{equation}
where $L_{SO}=(2m\alpha)^{-1}$ is the spin-orbit length. Notice the relation $L_{SO}^2=D\tau_s$, which amounts to say that the Dyakonov-Perel relaxation time is the time over which electrons diffuse over a spin-orbit length. $\sigma_0^{sH}$ corresponds to the expression given in Ref. \onlinecite{sin04}, i.e. the Drude formula evaluated without vertex corrections, of the spin Hall conductivity
\begin{equation}
\label{eq3}
\sigma_0^{sH}=\frac{e}{8\pi} \frac{2\tau}{\tau_{DP}}.
\end{equation}
In the diffusive regime $\alpha p_F\tau \ll 1$, one has $\tau_{DP} \gg \tau$, whereas in the ballistic one $\alpha p_F \tau \gg1 $ $\tau_{DP} \sim \tau$.
Notice that, by using Eq.(\ref{eq3}) and the expression for $s_0$,  the spin current (\ref{eq2}) can also be written as
\begin{equation}
\label{eq2bis}
J^z_y= \frac{1}{2m\alpha}\frac{1}{\tau_s}\left( s^y -s_0\right).
\end{equation}
By identifying the y-polarized spin current flowing along the x direction
\begin{equation}
\label{eq4}
J^y_x=-D\partial_x s^y
\end{equation}
the diffusion equation becomes
\begin{equation}
\label{eq5}
\partial_t s^y+\partial_x J^y_x =-2m\alpha J^z_y=-\frac{1}{L_{SO}}J^z_y
\end{equation}
which is the continuity-like equation for $s^y$
showing how the torque term associated to $s^y$ is expressed in terms of $J^z_y$.

\subsection{Single-interface problem}
In this Section we consider a static time-independent situation.
The idea is to analyze  a single-interface problem in order to understand the supercell numerical calculation. We then assume the following expression for the Rashba coefficient
\begin{equation}
\label{eq6}
\alpha (x)=\theta (x)\alpha_++\theta (-x)\alpha_-.
\end{equation}
Clearly at $\pm \infty$, one recovers the uniform case with complete cancellation of the spin current for the Rashba model with couplings $\alpha_{\pm}$. The strategy is to solve the diffusion equation in the two regions and connect them via the appropriate boundary conditions at $x=0$.
We seek a solution of the form
\begin{eqnarray}
\label{eq7}
s^y(x)&=&\theta (x)\left( s_{0,+}+\delta s_+ e^{-x/L_+}\right) \nonumber \\
&+& \theta (-x)\left( s_{0,-}+\delta s_- e^{x/L_+}\right).
\end{eqnarray}
In the above $s_{0,\pm}$ are the asymptotic values of the y-spin polarization at $\pm\infty$. Also $L_{\pm}$ are the corresponding spin-orbit lengths in the two regions.
The z-polarized spin current in the two regions reads
\begin{eqnarray}
\label{eq8}
J^z_{y,\pm}(x)&=&\sigma^{sH}_{0,\pm}E_x+\frac{D_{\pm}}{L_{\pm}} 
\left( s_{0,\pm}+
\delta s_{\pm} e^{\mp x/L_{\pm}}\right) \nonumber \\
&=&
\frac{D_{\pm}}{L_{\pm}} \delta s_{\pm} e^{\mp x/L_{\pm}},
\end{eqnarray}
where in the last step we used the fact that the constant terms cancel in each region. $D_{\pm}$ may differ in the two region because via the Fermi velocity they depend on the electron density.
Eq.(\ref{eq8})  shows that in the interface region there can be a spin current different form zero.
To evaluate it we need to know the two values $\delta s_{\pm}$. To this end we use the continuity of the spin density $s^y$ and of the spin current $J^y_x$ at the interface.
Continuity of the spin density gives
\begin{equation}
\label{continuity}
\delta s_--\delta s_+ =\Delta s_0,
\end{equation}
where $\Delta s_0=s_{0,+}-s_{0,-}=-eN_0 \tau (\alpha_+-\alpha_-)E_x$.
Continuity of the spin current instead gives
\begin{equation}
\label{continuity_current}
\frac{D_-}{L_-}\delta s_-+
\frac{D_+}{L_+}\delta s_+=0.
\end{equation}
After solving the system for $\delta s_-$ and $\delta s_+$,we get,  
\begin{eqnarray}
\delta s_-&=&\Delta s_0\frac{D_-}{L_-}\frac{1}{D_+L_+^{-1}+D_-L_-^{-1}}\label{eq9}\\
\delta s_+&=&-\Delta s_0
\frac{D_+}{L_+}\frac{1}{D_+L_+^{-1}+D_-L_-^{-1}}\label{eq10}.
\end{eqnarray}
From this it is clear that the spin Hall current averaged over the spin-orbit length
\begin{equation}
\label{eq11}
\frac{1}{L_-}\int_{-\infty}^0 {\rm d}x \ J^z_{y,-}+
\frac{1}{L_+}\int^{\infty}_0 {\rm d}x \ J^z_{y,+}=0,
\end{equation}
which is the analog of Eq.(\ref{eq5}). Furthermore the left-hand side correspond to the quantity $\gamma$ in the main text.
On the other hand we have that the total z-polarized spin current is different from zero
\begin{eqnarray}
\label{eq12}
&&\int_{-\infty}^0 {\rm d}x \ J^z_{y,-}+
\int^{\infty}_0 {\rm d}x \ J^z_{y,+}=\\
&&-\frac{2e N_0 D_+D_- m(\alpha_+-\alpha_-)^2\tau}{D_+L_+^{-1}+D_-L_-^{-1}} E_x
\equiv \sigma^{sH}L_{eff} E_x,  \nonumber 
\end{eqnarray}
where $L_{eff}$ is an effective length determined in terms of $L_{\pm}$ and $D_{\pm}$.
The spin Hall current is localized at the interface within a distance of order $L_{\pm}$.

The above calculation of a single interface suggests  that the calculation for the periodic modulation described in the main text can be analyzed in terms
of a series of interfaces. The spin current flows in the interface regions. However, by making the interface separation of the order of the spin relaxation length, one may have a spin current finite everywhere.

\subsection{Two-interface problem}
Here we consider the problem with two interfaces with  the model given by
\begin{equation}
\label{eq20}
\alpha (x)=\alpha_+\theta (l-|x|)+\alpha_- \theta (|x|-l).
\end{equation}
The solution for the spin density $s^y (x)$ is of the form
\begin{eqnarray}
s^y(x)&=&\theta (x-l) \left[s_{0,-}+\delta s_R e^{-(x-l)/L_-} \right]\nonumber\\
&+&\theta (-x-l) \left[s_{0,-}+\delta s_L e^{(x+l)/L_-} \right]\nonumber\\
&+&\theta (l-|x|)\left[s_{0,+}-\Delta s_0 \frac{\cosh (x/L_+)}{\cosh (l/L_+)} \right.\nonumber\\
&+&\left. \delta s_L \frac{\sinh ((l-x)/L_+)}{\sinh (2l/L_+)}+\delta s_R \frac{\sinh ((l+x)/L_+)}{\sinh (2l/L_+)}\right]
\nonumber
\end{eqnarray}
where the continuity of $s^y$ has already been implemented. The two constants $\delta s_R$ and $\delta s_L$ must be determined by imposing the conservation of the longitudinal y-polarized spin current $J^y_x$ as in the single-interface problem.

After going through  steps as for the single-interface we get
\begin{eqnarray}
\label{eq40}
&&\int_{-\infty}^{\infty} {\rm d}x \ J^z_y (x)= \\
&&-\frac{4e N_0 D_+D_- m(\alpha_+-\alpha_-)^2\tau \tanh (l/L_+)}{D_+L_+^{-1}\tanh (l/L_+)+D_-L_-^{-1}} E_x. \nonumber
\end{eqnarray}
The above equation has a simple interpretation in the limit
$l\rightarrow \infty$, when the two interfaces are far apart.
In this limit the two interfaces are independent. The total spin current is the sum of the spin currents flowing at the two interfaces. Eq.(\ref{eq40}) reduces indeed to twice the contribution (\ref{eq12}) of a single interface.
On the other hand when $l\sim L_+$ the two interfaces interact and the spin Hall current in non zero everywhere.
\vspace{1truecm}

\section{Continuity equations for the Rashba lattice model}
We provide here a discussion to clarify the role of an inhomogeneous Rashba SOC  to allow for 
a SHE in stationary conditions. The electron spin ${\bf S}$ obeys the equation of motion
\begin{equation}\label{eq:heisenberg}
\frac{d{\bf S}}{dt} = -i \lbrack {\bf S}, H\rbrack 
\end{equation}
and for the following it is convenient to separate the commutator into a term
related to the divergence of spin currents and a 'rest' ${\bf G}$

\begin{equation}
\lbrack {\bf S}, H\rbrack = -i div {\bf J} +i {\bf G} .
\end{equation}

As a result Eq. (\ref{eq:heisenberg}) can be interpreted in terms of a 
continuity equation
\begin{equation}\label{eq:galpha}
G^\alpha = div {\bf J}^\alpha + \frac{d S^\alpha}{d t}
\end{equation}
where ${\bf G}$ acts as 'source' term which in general is finite
due to the non-conservation of spin.

From evaluation of the commutators one finds the relation
\begin{equation}\label{eq:gj}
G_i^y=-\alpha_{i,i+y} J^z_{i,i+y} - \alpha_{i-y,i} J^z_{i-y,i}\,. ,
\end{equation}
i.e. a torque for the y-component of the spin is associated with
a z-spin current when the Rashba SOC $\alpha \ne 0$.
Upon combining Eq. (\ref{eq:gj}) with the y-component of Eq. (\ref{eq:galpha})
one finds 
\begin{equation}\label{eq:relation1}
\frac{d S_i^y}{d t}+[div {\bf J}^y]_i + \alpha_{i,i+y} J^z_{i,i+y} + \alpha_{i-y,i} J^z_{i-y,i} = 0
\end{equation}
and in particular for a system with periodic boundaries
\begin{equation}\label{eq:sumtot1}
\sum_i\left\lbrace\frac{d S_i^y}{d t}+ \alpha_{i,i+y} J^z_{i,i+y} + \alpha_{i-y,
i} J^z_{i-y,i}\right\rbrace = 0
\end{equation}
since the total divergence of any current has to vanish. For a homogeneous
SOC coupling Eq. (\ref{eq:sumtot1}) implies that the total z-spin current
has to vanish under stationary conditions.

\begin{figure}[htb]
\includegraphics[width=8cm,clip=true]{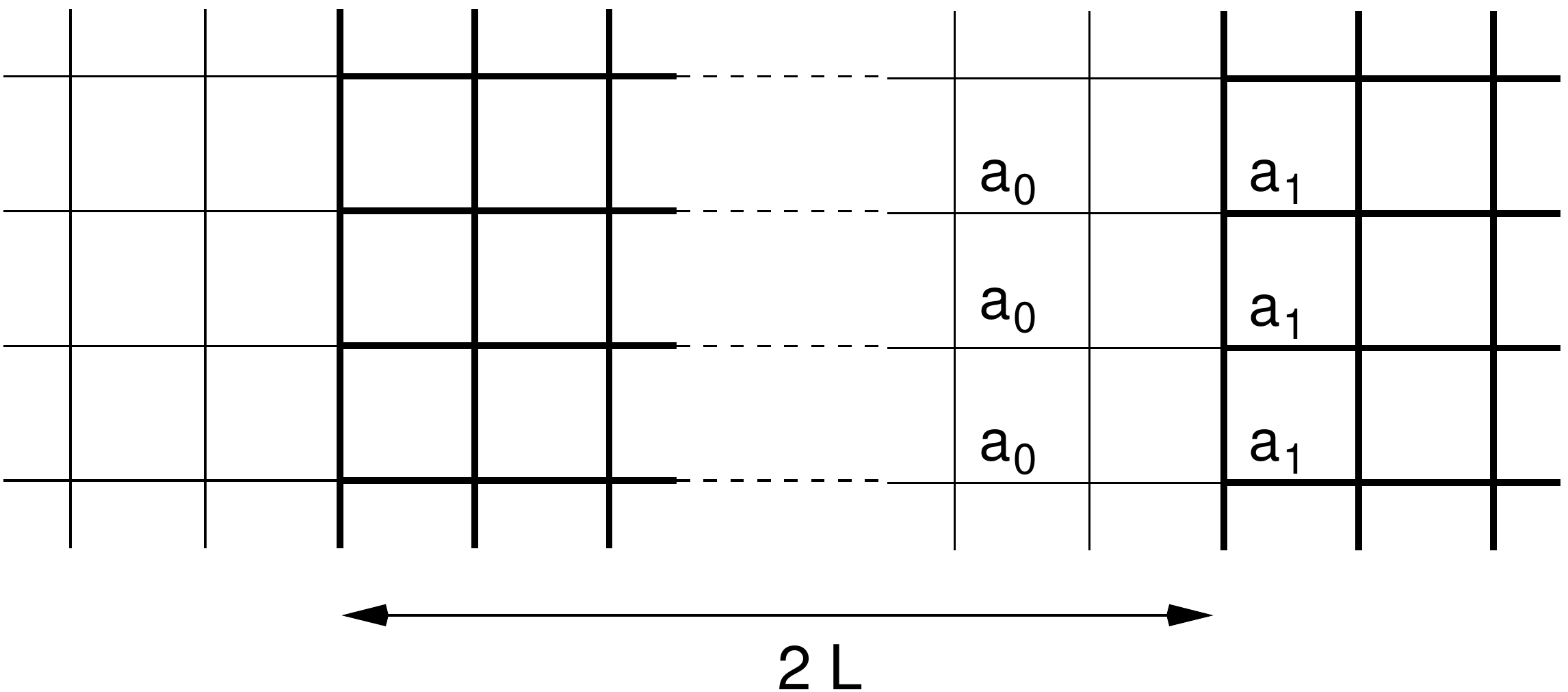}
\caption{Stripe-like modulation of the Rashba SOC  along the x-direction.
Stripes of width $L$ and coupling $a_0$ alternate with stripes of
the same width and coupling $a_1$ as indicated by thin and thick bonds,
respectively.}
\label{fig0}
\end{figure}

\begin{figure}[htb]
\includegraphics[width=8cm,clip=true]{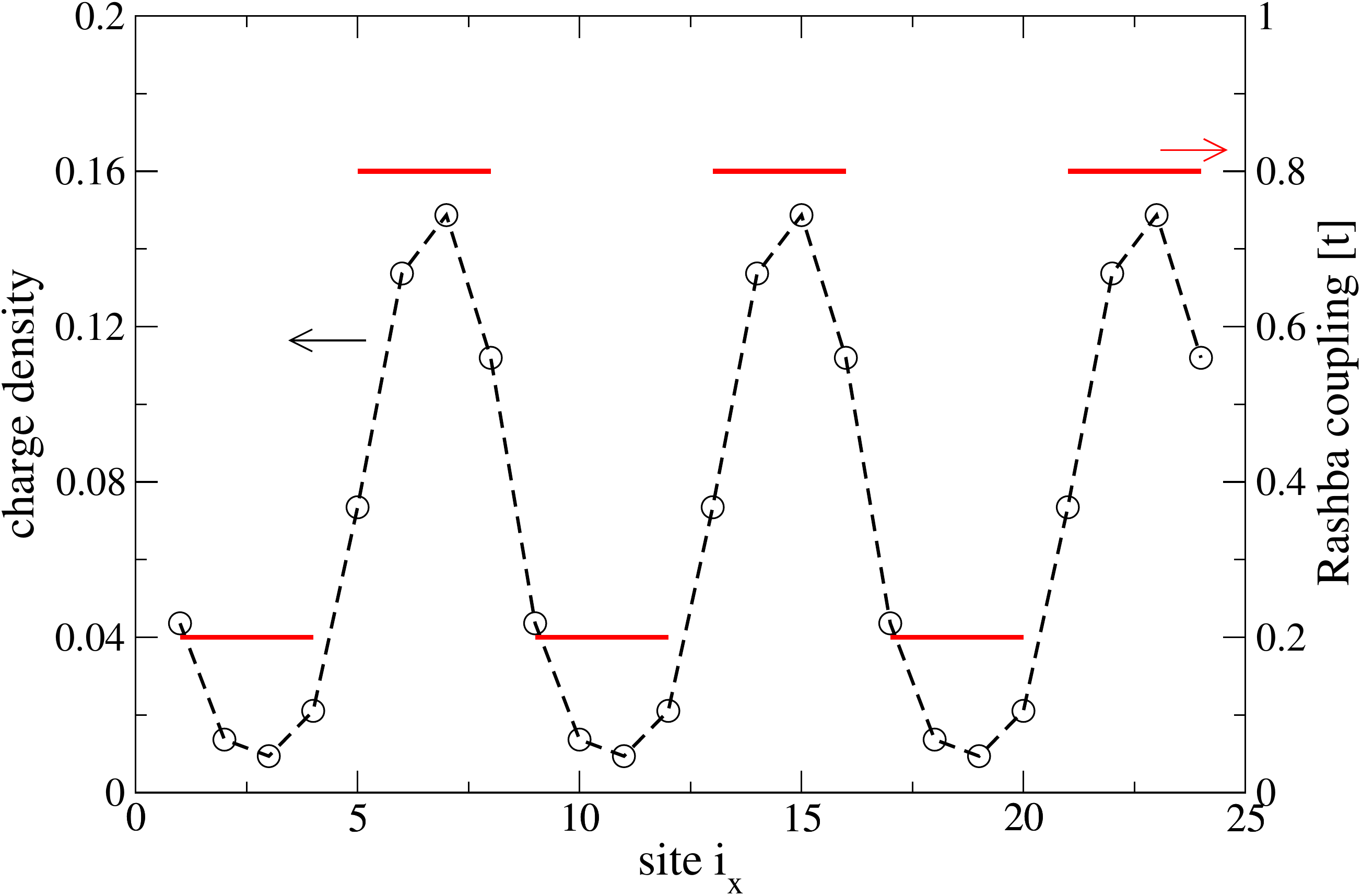}
\caption{Modulation of the coupling constant $\alpha_{i,i+x}$ (red) 
and charge density along the x-direction for the $L=4$ stripe-like Rashba SOC.
Particle concentration: $n=0.07$.}
\label{fig1}
\end{figure}

We exemplify the situation for a inhomogeneous Rashba SOC
which varies along the x-direction as
\begin{eqnarray}
\alpha_{i,i+x} &=& \frac{1}{2}\left\lbrack a_0+a_1+(a_0-a_1)\sgn(\sin\frac{2\pi i_x}{2 L})\right\rbrack \\
\alpha_{i,i+y} &=& \alpha_{i,i+x} \, ,
\end{eqnarray} 
i.e. one has stripes of width $L$ with Rashba SOC  $a_0$ alternating
with $L$-wide stripes having coupling $a_1$ (cf. Fig. \ref{fig0}). 
The inhomogeneity in $\alpha_{i,i+x}$ leads to a concomitant charge
modulation which is shown in Fig. \ref{fig1} for the case $L=4$.
Clearly the charge accumulates in regions with large Rashba SOC leading
to a CDW profile. We can therefore view this
model as an 'effective' model for a density dependent Rashba SOC.

\begin{figure}[hhh]
\includegraphics[width=7.5cm,clip=true]{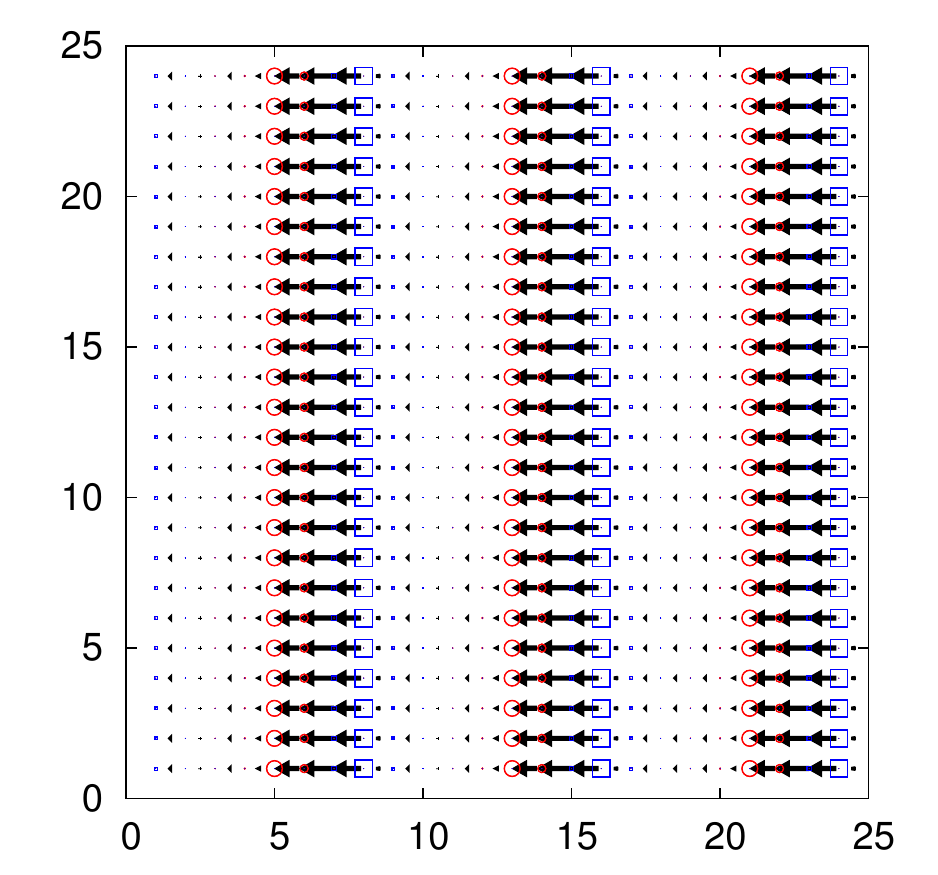}
\includegraphics[width=7.5cm,clip=true]{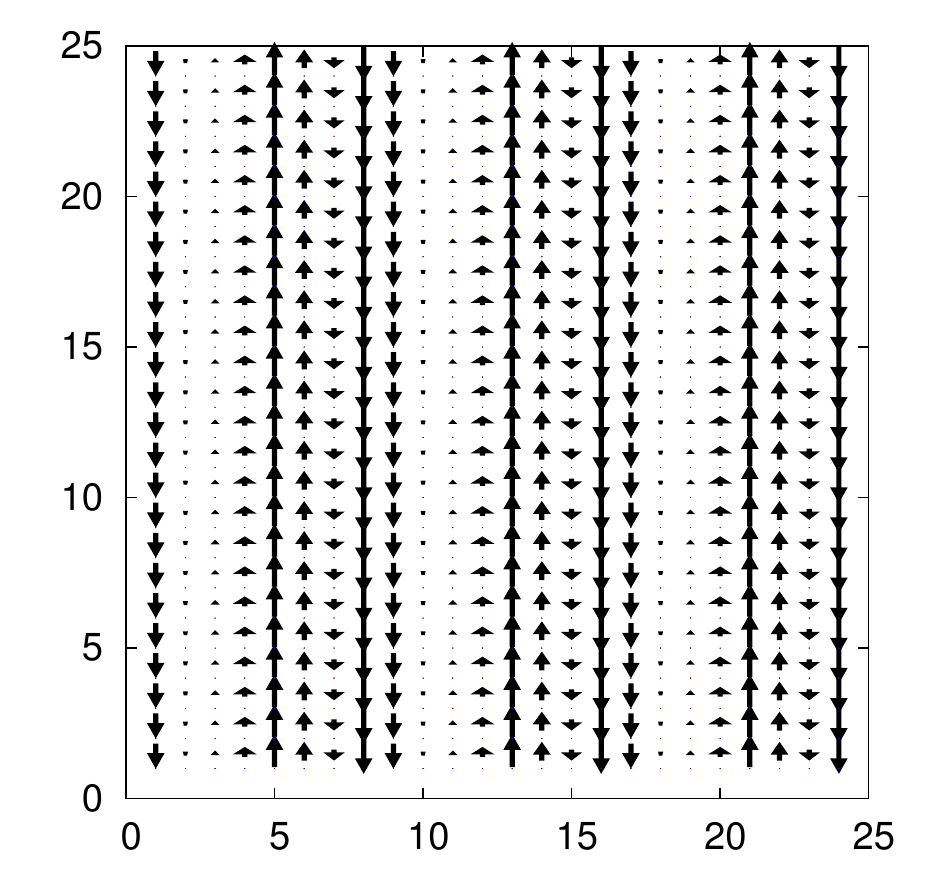}
\caption{Spin currents (arrows) and torques (squares) in the ground state
for a system with alternating $L=4$ stripes with Rashba couplings
$a_0=0.2$ and $a_1=0.8$, respectively. Top panel: y-components;
Bottom panel: z-components. Particle concentration: $n=0.07$.}
\label{fig2}
\end{figure}

Fig. \ref{fig2} shows the currents flowing in the (stationary) ground state.
Thus from Eq. (\ref{eq:galpha}) the torques $G_i^y$ are completely
determined by the divergence of the $J^y$ spin currents which are
shown by squares in the top panel of Fig. \ref{fig2}.   
As a consequence of Eq. (\ref{eq:gj})  a large z-spin current is flowing
on sites where also the $y$-torque is large in contrast to a
homogeneous system where $J^z=0$.
In the ground state the total y-torque $\sum_i G_i^y$ vanishes
so that from Eq. (\ref{eq:gj})
one obtains 
\begin{equation}\label{eq:gj2}
0=\sum_i\left\lbrack\alpha_{i,i+y} J^z_{i,i+y} + \alpha_{i-y,i} J^z_{i-y,i}
\right\rbrack\,. 
\end{equation}

The total number of $a_{0,1}$-stripes for a $N_x\times N_y$ lattice is
$n_{str}=N_x/(2L)$. Denote with $J^z_{0,1}$ the total z-spin current
flowing along the bonds of the $a_{0,1}$-stripes.
Then we can rewrite Eq. (\ref{eq:gj2}) as
\begin{equation}    
0=a_0 J^{z}_0 + a_1 J^z_1 \,\,\longrightarrow\,\, J^z_1=-\frac{a_0}{a_1}J^z_0 
\end{equation}
and the total z-spin current of the system is thus given by
\begin{equation}\label{eq:gs}
J^z_{tot}=n_{str}\left(J^{z}_0+J^{z}_1\right)=n_{str}J^{z}_0\left(1-\frac{a_0}{a_1}\right) \,.
\end{equation}
Now we can draw the following conclusions: The modulated Rashba SOC 
causes local torques $G^y_i$ which 
are related to local flows of z-spin currents. Provided that the total
y-torque vanishes the system thus exhibits a net flow of
$J^z_{tot}$ for $a_0 \neq a_1$ in the ground state.

The same analysis can now be applied in the presence of an electric
field  $E_x=-\partial_t A_x$ which couples to the system via
the charge current, i.e. 
$H'=-e\sum_i j^{ch}_{i,i+x} A_x(i,t)$ and $j^{ch}_{i,i+x}$.
Obviously Eq. (\ref{eq:relation1}) holds also in the
resulting non-equilibrium situation which we evaluate
in linear response. Each operator $\hat{O}_i$ in Eq. (\ref{eq:relation1})
reacts to the electric field according to 
$\hat{O}_i(\omega)=ie\Lambda_{ij}(\omega)E_j(\omega)/\omega$
and the correlation function is given by
\begin{displaymath}
\Lambda_{ij}(\omega) = -\frac{i}{N}\int_{-\infty}^\infty dt
\Theta(t-t') e^{i\omega(t-t')}
\langle \left[\hat{O}_i(t), j^{ch}_{j,j+x}(t')\right]\rangle \,.
\end{displaymath}

\begin{figure}[htb]
\includegraphics[width=8cm,clip=true]{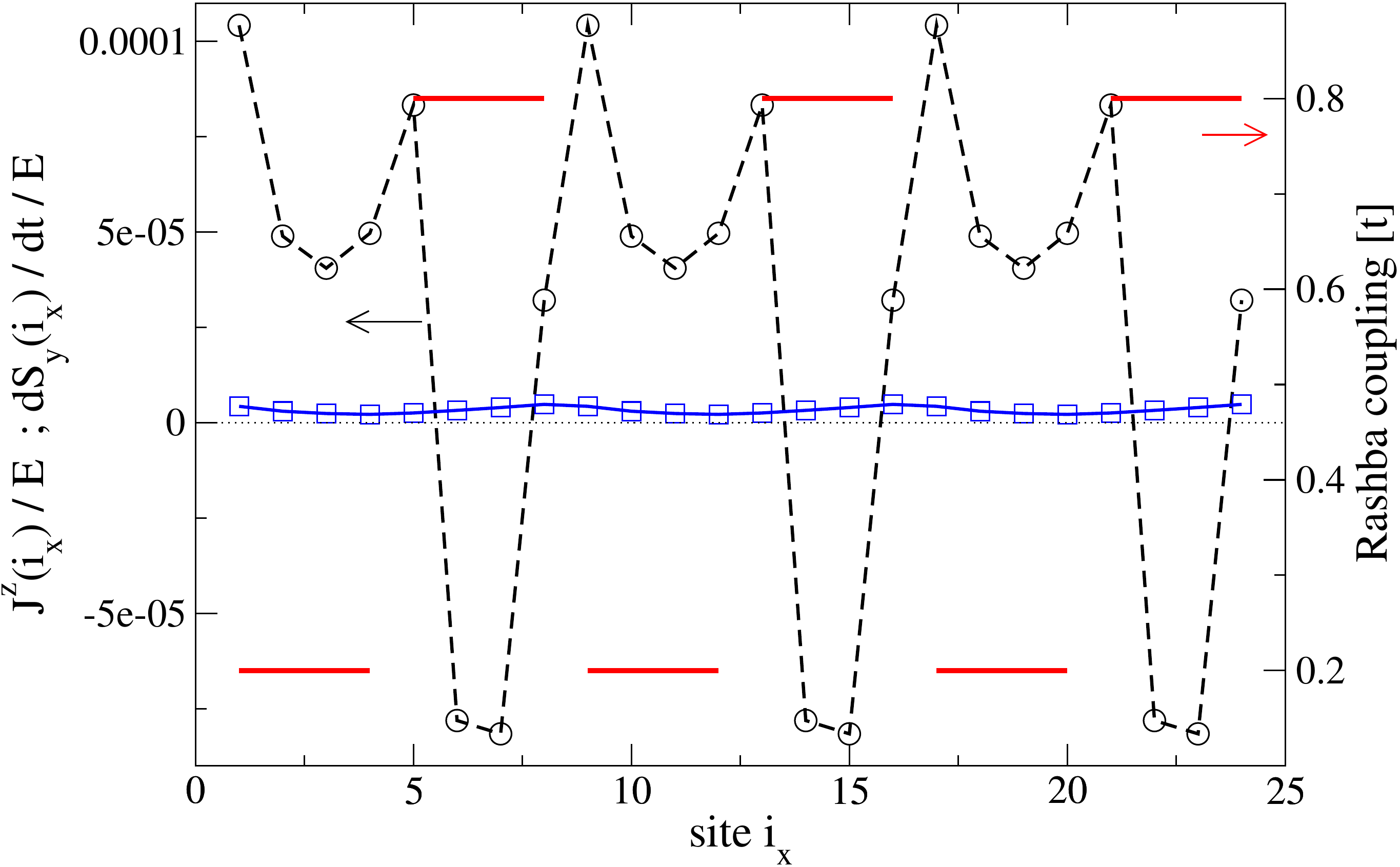}
\includegraphics[width=8cm,clip=true]{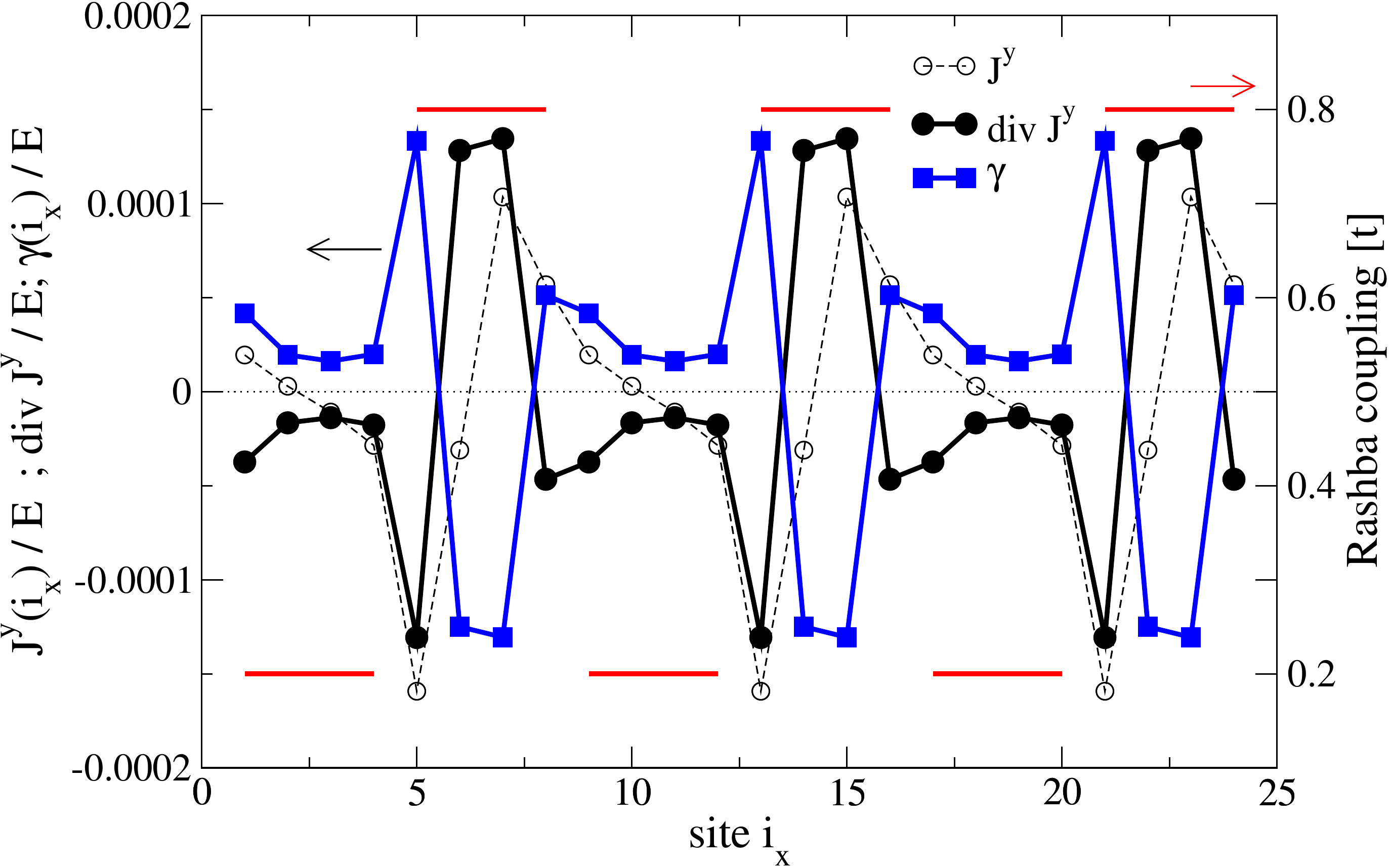}
\caption{Top panel: Modulation of the coupling constant $\alpha_{i,i+x}$ (red) 
and induced $J^z$-current (black, circles) for a cut along the x-direction 
and $L=4$ stripe-like SOC. The blue line (squares) shows the temporal 
change of the y-spin component $dS_y(i_x)/dt$. 
Bottom  panel: Induced $J^y$ spin current (black, open circles) and corresponding
divergence (black, full circles). The blue line (squares) shows the response
of $\gamma_{ix}=2\alpha_{i,i+y} J^z_{i,i+y}$ along the same cut.
Particle concentration: $n=0.07$.}
\label{sphedel}
\end{figure}

The top panel of Fig. \ref{sphedel} shows the induced $J^z$ spin current
together with the induced temporal change of $dS_y(i_x)/dt$ 
along a cut in x-direction of a $L=4$ striped system.
The dominant contribution to $J^z$ comes from the boundary regions
between small and large $\alpha$ stripes which gives rise to a
finite spin Hall conductivity.
Moreover, in contrast to the homogeneous case, where the induced
$J^z$ current and $dS_y(i_x)/dt$ are
equal 
\begin{equation}
\label{dimitrova1}
\frac{\partial S^y}{\partial t}=-\frac{2m \alpha}{\hbar} J^z_y.
\end{equation}
(although with opposite sign), we now
observe a much more stationary behavior. 
The reason can be deduced from the bottom panel which reports the x-dependence
of the induced $J^y$ spin current along with its divergence,
i.e. $[div J^y]_i\equiv J^y_{ix,ix+1}(i_x)-J^y_{ix-1,ix}$.
It turns out that the spatial behavior of the contribution 
$\gamma_{ix}=2\alpha_{i,i+y} J^z_{i,i+y}(i_x)$ is similar to $[div J^y]_i$
but opposite in sign which from
 Eq. (\ref{eq:relation1}) is responsible for the small value of
$dS_y(i_x)/dt$. 

From the above considerations we see that in a homogeneous Rashba
system a finite spin Hall conductivity necessarily implies
a non-stationary situation with the local accumulation of 
$S^y$ spin density. On the other hand an inhomogeneous
Rashba coupling partially shifts this time dependence to
a finite divergence of the $J^y$ spin currents which is
stationary due to non-conservation of spin.

\end{appendix}

\end{document}